\documentclass{article}

\usepackage{fullpage}
\usepackage{lmodern}
\usepackage[utf8]{inputenc}
\usepackage[T1]{fontenc}
\usepackage[USenglish]{babel}

\usepackage{amsmath, amssymb, amsthm}

\usepackage{booktabs}

\usepackage{algorithm}
\usepackage[noend]{algpseudocode}

\algrenewtext{While}[1]{\algorithmicwhile\ #1\textbf{:}}
\algrenewtext{For}[1]{\algorithmicfor\ #1\textbf{:}}
\algrenewtext{ForAll}[1]{\algorithmicforall\ #1\textbf{:}}
\algrenewtext{If}[1]{\algorithmicif\ #1\textbf{:}}
\algrenewcommand\algorithmicrequire{\textbf{Input:}}
\algrenewcommand\algorithmicensure{\textbf{Output:}}

\algnewcommand{\InlineFor}[1]{\State \algorithmicfor\ #1\textbf{:}}
\algnewcommand{\InlineForAll}[1]{\State \algorithmicforall\ #1\textbf{:}}
\algnewcommand{\InlineIf}[1]{\State \algorithmicif\ #1\textbf{:}}
\algnewcommand{\InlineElse}[1]{\State \algorithmicelse\textbf{:}}
\algnewcommand{\algorithmiclet}{\textbf{let}}
\algnewcommand{\Let}{\algorithmiclet\ }

\algnewcommand{\algorithmicoutput}{\textbf{output}}
\algnewcommand{\Output}{\algorithmicoutput\ }

\algnewcommand{\algorithmiccontinue}{\textbf{continue}}
\algnewcommand{\Continue}{\algorithmiccontinue\ }

\usepackage{enumitem}

\usepackage[round]{natbib}
\usepackage[%
  pdftitle={Beware of the Classical Benchmark Instances for the Traveling Salesman Problem with Time Windows},%
  pdfauthor={Francisco J.\ Soulignac},%
  colorlinks=true,%
  linkcolor=blue,%
  citecolor=blue,%
  urlcolor=blue]{hyperref}
\usepackage{doi}

\usepackage{tikz}
\usetikzlibrary{graphs, quotes, graphs.standard}
\usepackage{float}
\usepackage{caption}
\usepackage{subcaption}

\newcommand{\TSPTWM}{\mbox{TSPTW-M}}
\newcommand{\TSPTWD}{\mbox{TSPTW-D}}
\newcommand{\TSPTWTT}{\mbox{TSPTW-TT}}
\newcommand{\Network}{D}

\newcommand{\CustCount}{n}
\newcommand{\Vertex}{v}
\newcommand{\Wertex}{w}
\newcommand{\Zertex}{z}
\newcommand{\Release}{a}
\newcommand{\Deadline}{b}

\newcommand{\TravelTime}{\tau}
\newcommand{\ArrivalTime}{\delta}
\newcommand{\ArrivalTimeM}{\ArrivalTime_{\rm m}}

\newcommand{\DepartureTime}{\ArrivalTime^{-1}}
\newcommand{\DepartureTimeM}{\DepartureTime_{\rm m}}
\newcommand{\Horizon}{T}
\newcommand{\Route}{R}
\newcommand{\OtherRoute}{Q}
\newcommand{\Time}{t}
\newcommand{\Range}[1]{[\![#1]\!]}
\newcommand{\Duration}{\Delta}
\newcommand{\UpperBound}{ub}

\newcommand{\Sequence}[1]{\langle#1\rangle}
\newcommand{\RouteSet}{\mathcal{R}}
\newcommand{\Tree}{\mathcal{T}}
\newcommand{\VertexSet}{V}
\newcommand{\Unreachable}{U}
\newcommand{\Label}{\ell}
\newcommand{\TWWidth}{\omega}
\newcommand{\TWTightness}{\beta}
\newcommand{\TimeStart}{p}
\newcommand{\TimeEnd}{q}
\newcommand{\SpaceSize}{\sigma}
\newcommand{\Ascheuer}{\textsc{Asc}}
\newcommand{\sAscheuer}{\textsc{Asc-s}}
\newcommand{\lAscheuer}{\textsc{Asc-l}}
\newcommand{\DaSilvaUrrutia}{\textsc{DaS}}
\newcommand{\Dumas}{\textsc{Dum}}
\newcommand{\Gendreau}{\textsc{Gen}}
\newcommand{\Langevin}{\textsc{Lan}}
\newcommand{\OhlmannThomas}{\textsc{Ohl}}
\newcommand{\Pesant}{\textsc{Pes}}
\newcommand{\PotvinBengio}{\textsc{Pot}}
\newcommand{\DOhlmannThomas}{\textsc{Ohl-D}}

\newcommand{\Rifki}{\textsc{Rif}}
\newcommand{\LeraRomero}{\textsc{Ler22}}
\newcommand{\Tilk}{\textsc{Til22}}
\newcommand{\Rudich}{\textsc{Rud23}}
\newcommand{\Fontaine}{\textsc{Fon24}}

\DeclareMathOperator{\EAT}{EAT}
\DeclareMathOperator{\LDT}{LDT}
\DeclareMathOperator{\Best}{best}
\DeclareMathOperator{\Last}{last}

\newtheorem{proposition}{Proposition}

\newcommand{\CPP}{C\nolinebreak\hspace{-.05em}\raisebox{.2ex}{\small\bf +}\nolinebreak\hspace{-.10em}\raisebox{.2ex}{\small\bf +}20}

\title{Beware of the Classical Benchmark Instances for the Traveling Salesman Problem with Time Windows\thanks{Computers \& Operations Research, \url{https://doi.org/10.1016/j.cor.2026.107461}\\\copyright{} 2026. This manuscript version is made available under the CC-BY-NC-ND 4.0 licence \url{https://creativecommons.org/licences/by-nc-nd/4.0/}}}

\author{Francisco J.\ Soulignac\\\normalsize \texttt{francisco.soulignac@unq.edu.ar}}

\date{\normalsize Universidad Nacional de Quilmes. Departamento de Ciencia y Tecnología. Bernal, Buenos Aires, Argentina.\\ CONICET-Universidad de Buenos Aires. Instituto de Investigación en Ciencias de la Computación (ICC). Buenos Aires, Argentina.}

\begin{document}

\maketitle

\begin{abstract}

We propose a simple and exact method for the Traveling Salesman Problem with Time Windows and Makespan objective (\TSPTWM{}) that solves all instances of the classical benchmark with $50$ or more customers in less than ten seconds each.  Applying this algorithm as an off-the-shelf method, we also solve all but one of these instances for the Duration objective.  Our main conclusion is that these instances alone are no longer representative for evaluating the \TSPTWM{} and its Duration variant: their structure can be exploited to yield results that seem outstanding at first glance.  Additionally, caution is advised when designing hard training sets for machine learning algorithms.

~

\end{abstract}

\section{Introduction}
\label{sec:introduction}

In a Traveling Salesman Problem with Time Windows (TSPTW), a vehicle must visit a set of customers within their predefined time windows.  The vehicle cannot arrive late to a customer, although it can arrive early and wait until the beginning of its time window.  The journey begins at a depot to which the vehicle must return after visiting all the customers.  Several variants of the TSPTW are defined, depending on the objective function.  Three of the most studied variants involve minimizing the following: the completion time of the route, i.e., the time at which the vehicle returns to the depot (\TSPTWM{}; the total duration of the route, defined as the time elapsed between departure from and return to the depot (\TSPTWD{}); and the total travel time, which accounts only for time spent traveling between locations and disregards waiting times (\TSPTWTT{}).  From here on, we use the acronym TSPTW to refer to one of these three variants.

Research on the TSPTW began over forty years ago with foundational papers, such as \cite{ChristofidesMingozziToth1981} and \cite{Baker1983} for the \TSPTWM{}, \cite{Savelsbergh1985} for the \TSPTWTT{}, and \cite{Savelsbergh1992} for the \TSPTWD{}. Since then, numerous algorithms employing various solving strategies have been proposed for these three problem variants.

For exact approaches, significant efforts have focused on developing  formulations suited for branch-and-bound and branch-and-cut methods \cite[e.g.][]{Baker1983,LangevinDesrochersDesrosiersGelinasSoumis1993,AscheuerFischettiGroetschel2001,KaraKocAltiparmakDengiz2013}, including time-expanded networks \cite[e.g.][]{DashGuenluekLodiTramontani2012,BolandHewittVuSavelsbergh2017}, dynamic programming (both pure and incomplete) \cite[e.g.][]{DumasDesrosiersGelinasSolomon1995,MingozziBiancoRicciardelli1997,BalasSimonetti2001}, and anytime informed search methods \cite[e.g.][]{FontaineDibangoyeSolnon2023,KuroiwaBeck2023}. Other methods include dynamic programming within column generation and Lagrangian relaxation \cite[e.g.][]{BaldacciMingozziRoberti2012,TilkIrnich2017,Lera-RomeroMirandaBrontSoulignac2022}, constraint programming \cite[e.g.][]{PesantGendreauPotvinRousseau1998,FocacciLodiMilano2002}, and more recently, (multivalued) decision diagrams \cite[e.g.][]{GillardCoppeSchausCire2021,RudichCappartRousseau2023,CoppeGillardSchaus2024}.

In terms of heuristic and metaheuristic approaches, in addition to the pioneering works by \cite{Savelsbergh1985} and \cite{Savelsbergh1992}, notable contributions include heuristic methods by \cite{GendreauHertzLaporteStan1998,Calvo2000,Helsgaun2017}, tabu search by \cite{CarltonBarnes1996}, compressed annealing by \cite{OhlmannThomas2007}, and ant colony optimization (ACO) along with its Beam-ACO extensions by \cite{FavarettoMorettiPellegrini2006,Lopez-IbanezBlum2010,Lopez-IbanezBlumOhlmannThomas2013}. Other significant approaches include general variable neighborhood search \cite[e.g.][]{daSilvaUrrutia2010,MladenovicTodosijevicUrosevic2013,AmgharCordeauGendron2019,YeBartoliniSchneider2024}, iterated maximum large neighborhood search by \cite{Pralet2023}, and more recently, machine learning-based solvers \cite[e.g.][]{CappartMoisanRousseauPremont-SchwarzCire2021,KoolHoofGromichoWelling2022,ZhengHeZhouJinLi2023,BiMaZhouSongCaoWuZhang2024,ChenGongChenLiuWangYuZhang2025,li2025lmasklearnsolveconstrained}, among others.

Despite the many techniques developed for the TSPTW, the performance of most algorithms is evaluated on a subset of the \emph{classical} benchmark instances compiled by \cite{Lopez-IbanezBlum2023}, regardless of the objective function under consideration.  The common reasoning is that the validity of an instance depends more on the space of feasible routes than on the objective function.  Many machine learning algorithms replicate the generation process of some of these instances to create the large datasets of ``hard'' instances needed for training purposes~\cite[see e.g.][]{CappartMoisanRousseauPremont-SchwarzCire2021,KoolHoofGromichoWelling2022,BiMaZhouSongCaoWuZhang2024,li2025lmasklearnsolveconstrained}.  A recent exception to this rule for evaluating the TSPTW is \cite{Fontaine2024}, who designs hard instances with few customers based on a benchmark by \cite{RifkiChiabautSolnon2020}.

From the perspective of exact solvers, the classical benchmark is nearing the end of its useful life. \cite{BaldacciMingozziRoberti2012} solved all but one of the instances for the \TSPTWTT{} in at most one hour; \cite{RudichCappartRousseau2023,FontaineDibangoyeSolnon2023} jointly solved all but five instances for the \TSPTWM{} in at most one hour; and \cite{Lera-RomeroMirandaBrontSoulignac2022} solved all but fourteen of the considered instances for the \TSPTWD{} in at most three hours.  A natural question then arises: how will we develop the next generation of benchmarking instances? This question was addressed by \cite{daSilvaUrrutia2010}, who created a benchmark of large instances with $400$ customers, using similar methods to those employed for the classical instances.  Interestingly, \cite{FontaineDibangoyeSolnon2023} solved the benchmark by \cite{daSilvaUrrutia2010} in a few minutes, even though it is unable to solve some instances by \cite{Fontaine2024,RifkiChiabautSolnon2020} with $30$ customers within an hour.

The idea that the hardest instances for the TSPTW are those with loose time windows is widely accepted in the literature \cite[see e.g.][]{DumasDesrosiersGelinasSolomon1995}.  What sets the benchmark proposed by \cite{Fontaine2024} apart is its use of a parameter, $\TWTightness$, to systematically widen the time windows, as suggested by \cite{AriglianoGhianiGriecoGuerrieroPlana2019} for time-dependent problems. In a recent study, \cite{RifkiSolnon2025} investigate the effects of $\TWTightness$ and another parameter, $\alpha$, on the feasibility and hardness of Euclidean instances for the \TSPTWM{}.  Their main finding regarding feasibility is the presence of a pronounced phase transition between the infeasible and feasible regions.  On the topic of hardness, they empirically demonstrate that the exact method proposed by \cite{FontaineDibangoyeSolnon2023} benefits from tight time windows.  As the authors note, their analysis is crucial for the development of next-generation of benchmark instances and for training machine learning algorithms that requires datasets with varying levels of difficulty.  More recently, \cite{ChenGongChenLiuWangYuZhang2025} have questioned the time window generation processes used in previous machine learning studies, suggesting that they result in weak instances that are too easy to solve.

\subsection{Our contributions}

The main purpose of this article is to reinforce the conclusions of \cite{RifkiSolnon2025} and \cite{ChenGongChenLiuWangYuZhang2025}, demonstrating that the classical benchmark instances with $50$ or more customers, as well as the benchmark instances by \cite{daSilvaUrrutia2010}, should no longer be used alone to evaluate the \TSPTWM{}.  This is because these instances can be solved by a simple exact algorithm in less than ten seconds each.  Indeed, any method that includes a variant of our solver as a preprocessing step could yield misleading results that appear outstanding at first glance, regardless of how well the algorithm performs on alternative benchmarks.  Our method works by repeatedly running a simple informed search in a backward direction, prioritizing partial routes that reach the end depot earlier.  Despite its simplicity, it is surprising that the algorithm can solve all the large instances, including the four that remained open in previous studies.  However, the algorithm has serious limitations: it fails to solve some classical benchmark instances with fewer than $50$ customers and cannot solve any instance in the benchmark by \cite{Fontaine2024,RifkiChiabautSolnon2020} with 30 or 40 customers and loose time windows.

Regarding the \TSPTWD{}, we show that using our solver to minimize makespan as an off-the-shelf method is sufficient to solve all but one of the classical benchmark instances with $50$ or more customers, including the fourteen instances that remained unsolved in the study by \cite{Lera-RomeroMirandaBrontSoulignac2022}.  This time, all instances are solved in less than $30$ minutes each.  These results further reinforce the notion that the classical benchmark instances are particularly easy for our informed search method and, therefore, they should be complemented with other instances when evaluating the \TSPTWD{}.

Although our main conclusion is restricted to the \TSPTWM{} and the \TSPTWD{}, we believe that caution is also required when evaluating the \TSPTWTT{}.  Moreover, we argue that our results should be considered when designing new benchmark sets or training datasets for machine learning methods.  Neglecting to do so could lead to biased conclusions about the efficiency and effectiveness of the solvers \citep{ChenGongChenLiuWangYuZhang2025}.  The ideas proposed by \cite{AriglianoGhianiGriecoGuerrieroPlana2019}, \cite{Fontaine2024}, and \cite{RifkiSolnon2025}, which use a time window tightness parameter $\TWTightness$ to widen the time windows, provide a solid foundation for generating harder benchmark instances--even with fewer customers.

The fact that changing $\TWTightness$ yields harder instances certainly helps researchers evaluate the efficiency of their methods, but it does not imply that the synthetic time windows generated are significant for real-life situations.  For example, \cite{Ascheuer1996} discuss a real-life problem where tight time windows are designed to ensure that jobs are completed promptly in an online setting.  The short question, whose answer remains unclear to us, is whether the classical benchmark instances are truly representative of real-world problems. An affirmative answer could suggest that our algorithm, while too simplistic as a general purpose approach, might still be valid as a preprocessing step for filtering out easy instances before applying more sophisticated techniques.

\section{Problem Statement}
\label{sec:problem statement}

Throughout this article, we write $[j] = [0,j]$, $\Range{i,j} = [i,j]\cap \mathbb{N}$, and $\Range{j} = \Range{0,j}$ for $i,j \in \mathbb{R}$. Consider a \emph{transport network} described by a complete digraph $\Network$ with vertex set $\Range{\CustCount+1}$.  Vertices $0$ and $\CustCount+1$ represent the \emph{start} and \emph{end depots}, respectively, whereas the vertices in $\Range{1,\CustCount}$ correspond to \emph{customers}.  Each arc $\Vertex \to \Wertex$ of $\Network$ has a \emph{travel time} $\TravelTime(\Vertex,\Wertex)$ representing the time required to travel from $\Vertex$ to $\Wertex$.  Additionally, each vertex $\Vertex \in \Range{\CustCount+1}$ has a nonempty \emph{time window} $[\Release(\Vertex), \Deadline(\Vertex)]$ in which $\Vertex$ must be visited.  We assume that $[\Release(\Vertex), \Deadline(\Vertex)] \subseteq [\Release(0), \Deadline(0)] = [\Release(\CustCount+1), \Deadline(\CustCount+1)] = [\Horizon]$ for $v \in \Range{1,\CustCount}$, where $\Horizon$ is the \emph{planning horizon}.

A \emph{partial route} is a nonempty sequence of distinct vertices in $\Network$.  A partial route $\Route = \Sequence{\Vertex_0, \ldots \Vertex_k}$ is a \emph{route} if $\Vertex_0 = 0$ and $\Vertex_k = k = \CustCount+1$, i.e.,  $\Route$ visits each customer in $\Network$ exactly once.

For $i \in \Range{k}$ and $\Time \in [\Horizon]$, we can compute the earliest arrival time to $\Vertex_i$ when the vehicle traverses $\Route$ departing from $\Vertex_0$ at time $\Time$ using the following recurrence that takes waiting times into account:
\begin{displaymath}
 \ArrivalTime(\Route, \Time, i) = \begin{cases}
         \max\{\Time, \Release(\Vertex_0)\} & \text{if } i = 0, \\
         \max\{\ArrivalTime(\Route, \Time, i-1) + \TravelTime(\Vertex_{i-1}, \Vertex_{i}), \Release(\Vertex_i)\} & \text{otherwise.}
        \end{cases}
\end{displaymath}
The partial route $\Route$ is \emph{feasible} for time $\Time$ if $\Time \geq \Release(\Vertex_0)$ and $\ArrivalTime(\Route, \Time, i) \leq \Deadline(\Vertex_i)$ for every $i \in \Range{k}$.  Let $\ArrivalTime(\Route, \Time) = \ArrivalTime(\Route, \Time, k)$ if $\Route$ is feasible, and $\ArrivalTime(\Route, \Time) = \infty$ otherwise.  We refer to $\ArrivalTime(\Route, \Time)$ as the \emph{(earliest feasible) arrival time} of $\Route$ \emph{departing at time} $\Time$.  The \emph{makespan} of $\Route$ is $\ArrivalTime(\Route) = \ArrivalTime(\Route, \Release(\Vertex_0))$, while its \emph{duration} is $\Delta(\Route) = \min\{\ArrivalTime(\Route, \Time) - \Time \mid \Time \in [\Horizon]\}$. The goal of the \TSPTWM{}  (resp.\ \TSPTWD{}) is to find a route with minimum makespan (resp.\ duration).  Observe that the makespan (resp.\ duration) of such \emph{optimum routes} is infinity when no route is feasible for time $0$.

Sometimes it is convenient to know the departure time of the vehicle to arrive the last node of a partial route $\Route$ at a given time $\Time \in [\Horizon]$. We let $\DepartureTime(\Route, \Time)$ be the maximum $\Time' \in [\Horizon]$ such that $\ArrivalTime(\Route, \Time') = \Time$ ($\Time' = -\infty$ if no such $\Time'$ exists).  We refer to $\DepartureTime(\Route, \Time)$ as the \emph{(latest feasible) departure time} of $\Route$ \emph{arriving at time $\Time$}.

\subsection{Benchmark instances}
\label{sec:problem statement:benchmarks}

In this article, we consider $10$ sets of benchmark instances, comprising a total of $1337$ instances. The first $7$ sets contain $467$ instances and correspond to the classical instances compiled by \cite{Lopez-IbanezBlum2023}.  As explained in Section~\ref{sec:introduction}, the efficiency of most algorithms solving TSPTWs is evaluated on a subset of these instances.  Following \cite{Fontaine2024}, we refer to the individual benchmarks as \Ascheuer{} \citep{Ascheuer1996}, \Dumas{} \citep{DumasDesrosiersGelinasSolomon1995}, \Gendreau{} \citep{GendreauHertzLaporteStan1998}, \Langevin{} \citep{LangevinDesrochersDesrosiersGelinasSoumis1993}, \OhlmannThomas{} \citep{OhlmannThomas2007}, \Pesant{} \citep{PesantGendreauPotvinRousseau1998}, and \PotvinBengio{} \citep{PotvinBengio1996}. The travel times in \OhlmannThomas{} are obtained by rounding the original distances to the nearest integer.  The eighth benchmark, called \DOhlmannThomas{}, contains the same $25$ instances as \OhlmannThomas{} but with travel times rounded to the to the nearest tenth. This benchmark was considered in the context of the \TSPTWD{}.  The ninth set, called \DaSilvaUrrutia{}, comprises the $125$ instances from \cite{daSilvaUrrutia2010}, while the last set, called \Rifki{}, contains the remaining $720$ instances from \cite{Fontaine2024}, based on a benchmark by \cite{RifkiChiabautSolnon2020}.

To provide some context for the results discussed in the subsequent sections, we briefly review the origins of these benchmark sets.  Our primary focus is on the time windows, as their distributions appear to be more significant than the number of customers \citep[see e.g.][]{RifkiSolnon2025}.  We also note that many training sets for machine learning methods are generated using procedures similar to those employed in the creation of \Langevin{}, \Dumas{}, \Gendreau{}, \OhlmannThomas{}, and \DaSilvaUrrutia{} \citep[e.g.][]{CappartMoisanRousseauPremont-SchwarzCire2021,KoolHoofGromichoWelling2022,BiMaZhouSongCaoWuZhang2024,li2025lmasklearnsolveconstrained}.

\begin{itemize}
    \item \Langevin{} instances distribute the locations of a depot and $\CustCount \in \{20, 40, 60\}$ customers uniformly in $[\SpaceSize] \times [\SpaceSize]$ with $\SpaceSize = 100$.  The Euclidean distance, rounded to the nearest tenth, is used as the travel time.  Time windows are set by computing a second nearest-neighbor route $\Route = \Sequence{\Vertex_1, \ldots, \Vertex_{\CustCount+1}}$, and assigning $\Release(\Vertex_i) = \ArrivalTime(\Route, 0, i) - \mathcal{U}[0,\TWWidth/2]$ and $\Deadline(\Vertex_i) = \ArrivalTime(\Route, 0, i) + \mathcal{U}[0,\TWWidth/2]$ for a width $\TWWidth \in \{40, 60, 80\}$.  These instances are now considered trivial for the \TSPTWM{}.

    \item \Dumas{} instances were created using the same procedure as \Langevin{}, but with $\SpaceSize = 50$ for $\CustCount \in \{20, 40, 60\}$ and $\TWWidth \in \{20, 40, 60, 80, 100\}$; $n = 80$ and $\TWWidth \in \{20, 40, 60, 80\}$; $n \in \{100,150\}$ and $\TWWidth \in \{20, 40, 60\}$; and $n = 200$ and $\TWWidth \in \{20,40\}$. Quoting \cite{DumasDesrosiersGelinasSolomon1995}, ``for narrow widths, [our algorithm] is less than exponential [\ldots,] a 250-node problem with $\ell = 20$ is easy to solve in less than 10 seconds.''

    \item \Gendreau{} extends the benchmark by \Dumas{} for $n \in \{20,40,60,80,100\}$ and $\ell \in \{60, 80, \ldots, 200\}$.  In their words, ``[Our algorithm] would appear to be slower [\ldots] on instances with narrow time windows.  As noted by Dumas et al., the running time of their exact algorithm increases exponentially with time window width. [We generated our instances] to test the performance [\ldots] on instances with wide time windows.''

    \item \OhlmannThomas{} also extends \Dumas{} for $n = 150$ and $\ell \in \{120, 140, 160\}$, and for $n = 200$ and $\ell \in \{120,140\}$.  Their justification is that their instances ``show a solution method’s ability to not only cope with wide time windows, but also with large numbers of customers.''

    \item \DaSilvaUrrutia{} was designed to ``overcome the limitations of the other test sets'' as \cite{daSilvaUrrutia2010} needed ``instances with more than 200 customers and wider time windows to benchmark and compare algorithms.''  Additionally, they believed that the method by \cite{DumasDesrosiersGelinasSolomon1995} was ``a biased way to generate instances since it does not reflect the real cases. Moreover, there is no challenge to build a feasible solution because the second-nearest neighbor tour already is a feasible solution.''  To address these concerns, they generate random routes to assign the time windows, with $\CustCount \in \{200,250,300,350,400\}$, $\SpaceSize = 100$, and $\TWWidth \in \{100,200,300,400,500\}$.

    \item \PotvinBengio{} and \Pesant{} instances were obtained by extracting the route of a vehicle from a solution to Solomon's RC2 instances for the vehicle routing problem with time windows.  They are considered hard but have at most 45 customers.

    \item \Ascheuer{} instances are derived from a real-world application that minimizes the unloaded travel time of a stacker crane in an online setting.  The number of customers ranges from $10$ to $231$, while the travel time windows are tight to ensure jobs are not delayed for too long.

    \item \cite{RifkiChiabautSolnon2020} designed a benchmark set for time-dependent routing problems with travel times computed from shortest paths in the road network of Lyon, using realistic traffic simulations built from real-world data. To create \Rifki{}, \cite{Fontaine2024} removed the time dependency from the instances with $\CustCount \in \{20,30,40\}$.  Regarding time windows, \cite{Fontaine2024} follows the model presented by \cite{AriglianoGhianiGriecoGuerrieroPlana2019} for time-dependent instances, setting $\Release(\Vertex_i) = (\TWTightness\ArrivalTime(\Route, 0, i) - 40)$ and $\Deadline(\Vertex_i) = \ArrivalTime(\Route, 0, i) + 40$ for a random route $\Route = \Sequence{\Vertex_1, \ldots, \Vertex_{\CustCount+1}}$ and $\TWTightness \in \{0,0.25,0.5,1\}$.  Here, $\TWTightness \in [1]$ represents the tightness of the time windows, from the loosest $0$ to the tightest $1$.
\end{itemize}

\section{A Simple Solver for the \TSPTWM{}}
\label{sec:solver}

The solver for the \TSPTWM{} builds upon the best-first search algorithm depicted in Algorithm~\ref{alg:bfs}.  Given $\Time_0, \UpperBound \in [\Horizon]$, Algorithm~\ref{alg:bfs} determines whether a route $\Route$ in the transport network $\Network$ is feasible and arrives earlier than $\UpperBound$ when departing at time $\Time_0$, i.e., whether $\ArrivalTime(\Route, \Time_0) \leq \UpperBound$.  If $\Time_0 = 0$, then the makespan of $\Route$ is less than $\UpperBound$.

For convenience,  for every partial route $\Route$ starting at vertex $\Vertex$, we define: $\ArrivalTimeM(\Route) = \ArrivalTime(\Route, \max\{\Time_0, \Release(\Vertex)\})$ as the earliest arrival time when the vehicle leaves the depot at time $\Time_0$,  and $\DepartureTimeM(\Route) = \DepartureTime(\Route, \UpperBound)$ as the latest departure time when the vehicle cannot arrive at the end depot after time $\UpperBound$.

In essence, Algorithm~\ref{alg:bfs} is an informed search procedure that traverses a tree of partial routes $\Tree$ in a backward direction.  The root of $\Tree$ is the partial route $\Sequence{\CustCount+1}$, which only visits the end depot, while the children of a partial route $\Route$ are all the partial routes $\Sequence{\Wertex} + \Route$ obtained by prepending a vertex $\Wertex \not\in \VertexSet(\Route)$ to $\Route$. Here, $\VertexSet(\Route)$ denotes the set of vertices set already visited by $\Route$.

Within Algorithm~\ref{alg:bfs}, $\RouteSet$ maintains the \emph{unprocessed} nodes of $\Tree$, which have already had their parent processed (Steps \ref{alg:bfs:init}~and~\ref{alg:bfs:extension}).  Since $\Tree$ contains $\Theta(n!)$ nodes, two techniques are employed in Step~\ref{alg:bfs:unreachable} to prevent the exploration of the entire tree.  First, a set of \emph{unreachable} vertices $\Unreachable(\Wertex, \DepartureTimeM(\Route))$ is computed for every partial route $\Route$ that starts at vertex $\Wertex$, as described in Section~\ref{sec:basic:preprocessing}.  By construction, if $\Unreachable(\Wertex, \DepartureTimeM(\Route)) \not\subseteq \VertexSet(\Route)$, then no route $\Route'$ with suffix $\Route$ and $\ArrivalTime(\Route') < \UpperBound$ exists, and therefore the subtree of $\Tree$ rooted at $\Route$ can be discarded.  Similarly, if $\ArrivalTimeM(\Route) \geq \UpperBound$, then $\ArrivalTimeM(\Route') \geq \UpperBound$ for every route $\Route'$ with suffix $\Route$, and so this subtree can also be pruned.  These pruning strategies ensure that any partial route $\Route$ obtained in Step~\ref{alg:bfs:stop} is feasible and has $\ArrivalTimeM(\Route) < \UpperBound$.

Following \cite{FontaineDibangoyeSolnon2023}, each iteration of the main loop (Steps~\ref{alg:bfs:main loop start}--\ref{alg:bfs:main loop end}) processes one route $\Route$ of length $k$, if one exists, for every $k \in \Range{n+1}$ (Steps~\ref{alg:bfs:k loop begin}--\ref{alg:bfs:k loop end}).  All extensions of $\Route$ in Step~\ref{alg:bfs:extension} will have length $k+1$, and thus can be processed within the same iteration of the main loop.  The goal is to compute the output route as early as possible.  To guide the search, a heuristic is applied to decide which partial route $\Route$ of length $k$ to process next (Step~\ref{alg:bfs:heuristic}).

Note that $\UpperBound - \ArrivalTimeM(\Route)$ represents an upper bound on the time that can still be saved by any route with suffix $\Route$.  In the extreme case where $\ArrivalTimeM(\Route) \geq \UpperBound$, no improvement is possible, and $\Route$ is discarded (Step~\ref{alg:bfs:unreachable}).  Maximizing $\UpperBound - \ArrivalTimeM(\Route)$ maximizes the opportunity for the vehicle to wait at previous customers, if necessary.   In event of ties, Step~\ref{alg:bfs:heuristic} minimizes $\DepartureTime(\Route, \ArrivalTimeM(\Route))$ to maximize the earliest departure time of $\Route$ that yields no waiting due to time windows. The goal is to allow the vehicle to wait at the previous (yet unknown) customers without significantly affecting the earliest arrival time.

Finally, Algorithm~\ref{alg:bfs} avoids extending a partial route $\Route$ if it can prove that it is $\UpperBound$-dominated.  Formally, $\Route$ is \emph{$\UpperBound$-dominated} by a partial route $\OtherRoute$ of length $|\Route|$ if, for every route $\Route'$ with suffix $\Route$, either the earliest arrival time of $\Route'$ is $\ArrivalTimeM(\Route') \geq \UpperBound$, or there exists a route $\OtherRoute'$ with suffix $\OtherRoute$ whose earliest arrival time is $\ArrivalTimeM(\OtherRoute') < \UpperBound$.  The extension of $\Route$ can be avoided when it is $\UpperBound$-dominated by an extended partial route $\OtherRoute \neq \Route$.  Algorithm~\ref{alg:bfs} applies the following rule to discard $\UpperBound$-dominated routes.

\begin{proposition}\label{dom-rule:basic}
 Let $\Route$ be a partial route from vertex $\Vertex$ to the end depot $\CustCount+1$ with $\DepartureTimeM(\Route) \geq \Time_0$.  If there exists a partial route $\OtherRoute \neq \Route$ from $\Vertex$ to $\CustCount+1$ such that its earliest arrival time is $\ArrivalTimeM(\OtherRoute) < \UpperBound$, it visits the vertices $\VertexSet(\OtherRoute) = \VertexSet(\Route)$, its latest departure is $\DepartureTimeM(\OtherRoute) \geq \DepartureTimeM(\Route)$ and one of the following holds:
 \begin{enumerate}
  \item its latest departure is $\DepartureTimeM(\OtherRoute) > \DepartureTimeM(\Route)$, or
  \item $\ArrivalTime(\Route, \DepartureTimeM(\Route)) = \UpperBound$ (i.e., $\Route$ cannot arrive earlier than $\UpperBound$ when departing as late as possible), or
  \item $\ArrivalTime(\OtherRoute, \DepartureTimeM(\OtherRoute)) < \UpperBound$ (i.e., $\OtherRoute$ arrives earlier than $\UpperBound$ when departing as late as possible).
 \end{enumerate}
 then $\Route$ is $\UpperBound$-dominated by $\OtherRoute$.
\end{proposition}

Steps~\ref{alg:bfs:dominance begin}--\ref{alg:bfs:dominance end} apply Proposition~\ref{dom-rule:basic} to determine if $\Route$ is $\UpperBound$-dominated. A table is maintained, indexed by a vertex set $\VertexSet$ and a vertex $\Vertex$, to store the greatest $\DepartureTimeM(\OtherRoute)$ among every processed route $\OtherRoute$ with $\VertexSet(\OtherRoute) = \VertexSet$ and first vertex $\Vertex$.  The table also stores $\ArrivalTime(\OtherRoute, \DepartureTimeM(\OtherRoute))$; note that $\ArrivalTimeM(\OtherRoute) < \UpperBound$ follows by invariant.  Because of the search heuristic, Algorithm~\ref{alg:bfs} can process an $\UpperBound$-dominated partial route before extending any of its dominators.  We note that Algorithm~\ref{alg:bfs} could discard all the routes minimizing the makespan because some of their prefixes are $\UpperBound$-dominated.  This explains why Algorithm~\ref{alg:bfs} is only a decision method rather than an optimization method, which justifies the decision to stop as soon as a route with $\ArrivalTimeM(\Route) < \UpperBound$ is found.

\begin{algorithm}[tbh]
\caption{Backward Best First Search Labeling}\label{alg:bfs}
\begin{algorithmic}[1]
    \Require a transport network $\Network$, a departure time $\Time_0 \in [\Horizon]$ and a latest arrival time $\UpperBound \in [\Horizon]$.
    \Ensure a route $\Route$ with $\ArrivalTimeM(\Route) < \UpperBound$ or a message that no such $\Route$ exists.

    \State \Let $\RouteSet = \{\Sequence{n+1}\}$ be a family of partial routes\label{alg:bfs:init}

    \While{$\Route$ is nonempty}\label{alg:bfs:main loop start}
        \ForAll{$k = 0, \ldots, \CustCount+1$}\label{alg:bfs:k loop begin}
            \InlineIf{$\RouteSet$ has no partial route of length $k$} \Continue

            \State Remove $\Route \in \RouteSet$ of length $k$ and minimum $\langle\ArrivalTimeM(\Route), \DepartureTime(\Route, \ArrivalTimeM(\Route))\rangle$\label{alg:bfs:heuristic}

            \InlineIf{$k = \CustCount+1$} \Output $\Route$ and halt\label{alg:bfs:stop}

            \State \Let $\Vertex$ be the first vertex in $\Route$ and $(d,a) := \Best(\VertexSet(R),\Vertex)$\label{alg:bfs:dominance begin}

            \InlineIf{$\DepartureTimeM(\Route) < d$} \algorithmiccontinue
            \InlineIf{$\DepartureTimeM(\Route) = d$ and $a < \UpperBound$} \algorithmiccontinue
            \InlineIf{$\DepartureTimeM(\Route) = d$ and $\ArrivalTime(\Route, \DepartureTimeM(\Route)) = \UpperBound$} \algorithmiccontinue;

            \State \Let $\Best(\VertexSet(\Route), \Vertex) = \langle\DepartureTimeM(\Route), \ArrivalTime(\Route, \DepartureTimeM(\Route))\rangle$\label{alg:bfs:dominance end}

            \ForAll{$\Wertex \in \Range{\CustCount} - \VertexSet(\Route)$}

                \State \Let $\Route' = \Sequence{\Wertex} + \Route$\label{alg:bfs:extension}

                \InlineIf{$\ArrivalTimeM(\Route') < \UpperBound$ and $\Unreachable(\Wertex, \DepartureTimeM(\Route')) \subseteq \VertexSet(\Route')$} insert $\Route'$ into $\RouteSet$ \label{alg:bfs:unreachable}\label{alg:bfs:main loop end}\label{alg:bfs:k loop end}
            \EndFor
        \EndFor
    \EndWhile
    \State \Output ``no route $\Route$ exists with $\ArrivalTimeM(\Route) < \UpperBound$''
\end{algorithmic}
\end{algorithm}

To solve the \TSPTWM{}, we repeatedly execute Algorithm~\ref{alg:bfs} as described in Algorithm~\ref{alg:basic}.  The output of this process is the route that minimizes the earliest arrival time, $\ArrivalTime(\Route, \Time_0)$, for a given departure time $\Time_0 \in [\Horizon]$.

\begin{algorithm}[tbh]
\caption{Basic solver for the \TSPTWM{}}\label{alg:basic}
\begin{algorithmic}[1]

    \Require a transport network $\Network$ and a departure time $\Time_0 \in [\Horizon]$

    \Ensure a route $\Route$ minimizing $\ArrivalTimeM(\Route) < \Horizon$ or a message no such route exists.

    \State compute the unreachable function for Step~\ref{alg:bfs:unreachable} as in Section~\ref{sec:basic:preprocessing}\label{alg:basic:preprocessing}

    \State \Let $\Route^* = \emptyset$ and $\Horizon' = \Horizon + 1$.

    \While{Algorithm~\ref{alg:bfs} with input $\Network$, $\Time_0$, and $\UpperBound = \Horizon'$ outputs a route $\Route$}

        \State \Let $\Horizon' = \ArrivalTimeM(\Route)$ and $\Route^* = \Route$

    \EndWhile

    \InlineIf{$\Route^* \neq \emptyset$} \Output $\Route^*$, \algorithmicelse{} \Output ``the desired route does not exist''.

\end{algorithmic}
\end{algorithm}

\paragraph{Implementation details}

Although our implementation of Algorithm~\ref{alg:bfs} is not particularly efficient, we have taken measures to improve its running time. Each partial route $\Route$ is implemented with a label $\ell(\Route)$ that uses $n + O(1)$ bits.  The label $\Label(\Route)$ is a tuple containing  the following information:
\begin{itemize}
 \item The first vertex $\Vertex$ of $\Route$,
 \item A bitset representing the set of visited vertices $\VertexSet(\Route)$,
 \item The latest departure time $\DepartureTimeM(\Route)$,
 \item The earliest arrival time $\ArrivalTimeM(\Route)$,
 \item The times $\ArrivalTime(\Route, \DepartureTimeM(\Route))$ and $\DepartureTime(\Route, \ArrivalTimeM(\Route))$, and
 \item A pointer to the label representing its parent in $\Tree$.
\end{itemize}
By traversing the branch of $\Tree$ from node $\Route$ using the parents pointers, we can obtain all the vertices in $\Route$ in $O(n)$ time.  Each time $\Route$ is extended into a partial route $\Route'$ in Step~\ref{alg:bfs:extension}, the times $\ArrivalTime(\Route', \DepartureTimeM(\Route'))$ and $\DepartureTime(\Route', \ArrivalTimeM(\Route'))$ are computed in $O(n)$ time.  Additionally, $\Unreachable(\Wertex, \DepartureTimeM(\Route'))$ is computed in $O(n)$ time, as discussed in Section~\ref{sec:basic:preprocessing}.

The family of routes $\RouteSet$ is implemented using a priority queue that holds all the unprocessed routes of length $k$, for every $k \in \Range{\CustCount+1}$.    Finally, we maintain a hash table $\Best[\Vertex]$ for each $\Vertex \in \Range{\CustCount+1}$, indexed by $\VertexSet(\Route)$, which is used in Steps~\ref{alg:bfs:dominance begin}~and~\ref{alg:bfs:dominance end}.

\subsection{The Unreachable Function}
\label{sec:basic:preprocessing}

As illustrated in Algorithm~\ref{alg:basic} (Step~\ref{alg:basic:preprocessing}), we apply a preprocessing step to compute an \emph{unreachable function} \mbox{$\Unreachable \colon \Range{n+1}\times [\Horizon] \to 2^{\Range{n+1}}$}, which is later used by Algorithm~\ref{alg:bfs} to prune some branches of the tree $\Tree$ of partial routes.  For each vertex $\Vertex \in \Range{n+1}$ and time $\Time \in [\Horizon]$, the function $\Unreachable(\Vertex, \Time)$ returns a set of vertices that cannot appear before $\Vertex$ in any feasible route $\Route$ that visits $\Vertex$ at a time $\Time' \leq \Time$.  By construction, $\Unreachable(\Vertex, \Time) \subseteq \Unreachable(\Vertex, \Time')$, so $\Unreachable$ is implemented by storing $\Unreachable(\Vertex, \Time)$ only for those times $\Time \in [\Horizon]$ where the function changes. As a result, each $\Unreachable(\Vertex,\bullet)$ has $O(n)$ values, which allows for queries to be answered in $O(n)$ time.

The computation of $\Unreachable$ follows these steps. First, we run an all-pairs shortest path algorithm twice to obtain the following functions:
\begin{align*}
 \EAT(\Wertex,\Vertex) &= \min\{\ArrivalTime(\Route, \Release(\Wertex)) \mid \Route \text{ is a feasible route from } \Wertex \text{ to }\Vertex\} \cup \{\infty\}\\
 \LDT(\Wertex,\Vertex) &= \max\{\DepartureTime(\Route, \Deadline(\Vertex)) \mid \Route \text{ is a feasible from } \Vertex \text{ to } \Wertex\}  \cup \{-\infty\}
\end{align*}
for every pair of vertices $\Vertex, \Wertex$ in $\Network$. Here, $\EAT(\Wertex, \Vertex)$ represents the earliest arrival time at $\Vertex$ after visiting $\Wertex$, and $\LDT(\Wertex, \Vertex)$ represents the latest departure time from $\Wertex$ before visiting $\Vertex$.

Next, we compute the precedence relation $\prec$, where $\Vertex \prec \Wertex$ if and only if $\EAT(\Wertex,\Vertex) = \infty$ or there exists a third vertex $\Zertex \in \Range{n+1} - \{\Vertex, \Wertex\}$ such that:
\begin{align*}
 \EAT(\Zertex, \Wertex) &> \LDT(\Wertex, \Vertex),\\
 \EAT(\Wertex, \Vertex) &> \LDT(\Vertex, \Zertex), \text{and} \\ \EAT(\Wertex, \Zertex) &> \LDT(\Zertex, \Vertex).
\end{align*}
This relation ensures that no feasible route can visit $\Wertex$ before $\Vertex$ when $\Vertex \prec \Wertex$. Therefore, for every $\Time \in \mathbb{R}$, we set $\Wertex \in \Unreachable(\Vertex, \Time)$ if $\Vertex \prec \Wertex$. Otherwise, we set $\Wertex \in \Unreachable(\Vertex, \Time)$ if and only if $\Time < \EAT(\Wertex, \Vertex)$.

Once the precedence relation $\prec$ is computed, we update the time windows by setting:
\begin{displaymath}
 \Release(\Wertex) := \max\{\Release(\Wertex), \EAT(\Vertex, \Wertex)\} \quad \text{and} \quad \Deadline(\Vertex) := \min\{\Deadline(\Vertex), \LDT(\Vertex, \Wertex)\}
\end{displaymath}
for every pair of vertices $\Vertex, \Wertex \in \Range{\CustCount+1}$ such that $\Vertex \prec \Wertex$. If any time window is modified, we recompute $\Unreachable$ and repeat the process until no further changes are made.

\subsection{Solving the integer \TSPTWD{}}

Recall that the \TSPTWD{} consist of finding a route $\Route$ with minimum duration $\Duration(\Route) = \min\{\ArrivalTime(\Route, \Time) - \Time \mid \Time \in [\Horizon]\}$ for a given transport network $\Network$.  Clearly, if the travel times and all the time windows in $\Network$ are integers, then a solution to the \TSPTWD{} can be obtained by running a solver for the \TSPTWM{} $O(\Horizon)$ times.  This is the approach Algorithm~\ref{alg:tsptw-d} uses to solve the \TSPTWD{}.

Specifically, Algorithm~\ref{alg:tsptw-d} maintains a sliding window $\Range{\TimeStart, \TimeEnd}$, a route $\Route$ that minimizes $\ArrivalTime(\Route, \TimeStart-1) = \TimeEnd$, and a route $\Route^*$ with a time $\Time^*$ that minimizes $\ArrivalTime(\Route^*, \Time^*) - \Time^*$ for $\Time^* \in \Range{\TimeStart-1}$.   Steps~\ref{alg:tsptw-d:init-r}--\ref{alg:tsptw-d:init-t} initialize these variables.  Step~\ref{alg:tsptw-d:init-t} also computes $\Last_\TimeStart$, which is the latest of the departure times of all the routes in $\Network$.  Thus, when $\TimeStart > \Last_\TimeStart$, $\Route^*$ is a route with minimum duration.

Throughout the main loop (Steps~\ref{alg:tsptw-d:loop start}--\ref{alg:tsptw-d:loop end}) the route $\Route$ is updated into a new route that minimizes $\ArrivalTime(\Route, \TimeStart) \geq \TimeEnd$.  Note that $\ArrivalTime(\Route, \TimeStart) = \TimeEnd$ when Step~\ref{alg:tsptw-d:tsptw-m} is skipped.  Then, $\TimeEnd$, $\Route^*$, and $\Time^*$ are updated accordingly in Steps~\ref{alg:tsptw-d:tsptw-m}, \ref{alg:tsptw-d:update}, and~\ref{alg:tsptw-d:increase p}.

\begin{algorithm}[tbh]
\caption{Sliding window solver for the integer \TSPTWD{}}\label{alg:tsptw-d}
\begin{algorithmic}[1]

    \Require a transport network $\Network$ in which $\TravelTime$ and the time windows are integer

    \Ensure a route $\Route^*$ and a time $\Time^*$ minimizing $\Duration(\Route^*) = \ArrivalTime(\Route,\Time^*) - \Time^*$ or a message that $\Network$ is infeasible.

    \State \Let $\Route$ be a route minimizing $\ArrivalTime(\Route)$\label{alg:tsptw-d:init-r}

    \State halt with \Output ``$\Network$ is infeasible'' \algorithmicif{} $\ArrivalTime(\Route) = \infty$

    \State heuristically find a route $\Route^*$ with $\ArrivalTime(\Route^*, \Time^*) = \ArrivalTime(\Route)$ that maximizes $\Time^* \in [\Horizon]$\label{alg:tsptw-d:init-r*}

    \State \Let $\TimeStart = \Time^* + 1$, $\TimeEnd = \ArrivalTime(\Route^*)$, and $\Last_\TimeStart = \max\{\DepartureTime(\Route, \Horizon) \mid \Route \text{ is a route in } \Network\}$\label{alg:tsptw-d:init-t}

    \While{$\TimeStart \leq \Last_\TimeStart$}\label{alg:tsptw-d:loop start}
        \State heuristically update $\Route$ to minimize $\ArrivalTime(\Route, \TimeStart)$\label{alg:tsptw-d:local search 1}
        \If{$\ArrivalTime(\Route, \TimeStart) > \TimeEnd$}
            \State \Let $\Route$ be a route that minimizes $\ArrivalTime(\Route, \TimeStart)$ and \Let $\TimeEnd =  \ArrivalTime(\Route, \TimeStart)$\label{alg:tsptw-d:tsptw-m}
        \EndIf
        \State heuristically update $\Route$ and $\TimeStart$ to maximize $\TimeStart$ with $\ArrivalTime(\Route,\TimeStart) = \TimeEnd$\label{alg:tsptw-d:local search 2}
        \InlineIf{$\ArrivalTime(\Route^*, \Time^*) - \Time^* > \ArrivalTime(\Route, \TimeStart) - \TimeStart$} \Let $\Route^* = \Route$ and $\Time^* = \TimeStart$\label{alg:tsptw-d:loop end}\label{alg:tsptw-d:update}
        \State \Let $\TimeStart = \TimeStart + 1$\label{alg:tsptw-d:increase p}
    \EndWhile
    \State \Output $\Route^*$ and $\Time^*$\label{alg:tsptw-d:output}
\end{algorithmic}
\end{algorithm}

Regarding the implementation of Algorithm~\ref{alg:tsptw-d}, we run Algorithm~\ref{alg:basic} with $\Time_0 = 0$ and $\UpperBound = \Horizon+1$ for Step~\ref{alg:tsptw-d:init-r}.  Similarly, we invoke Algorithm~\ref{alg:basic} with $\Time_0 = \TimeStart$ and $\UpperBound = \min\{\Last_\TimeEnd + \Duration(\Route^*), \Horizon+1\}$ for Step~\ref{alg:tsptw-d:tsptw-m}.  Let $\Network^{-1}$ be the \emph{reverse} transport network with vertex set $\Range{\CustCount+1}$, where the travel time of an arc $\Vertex \to \Wertex$ is $\TravelTime(\Wertex, \Vertex)$, and the time window of a vertex $\Vertex$ is $[\Horizon - \Deadline(\Vertex), \Horizon - \Release(\Vertex)]$.  In $\Network^{-1}$, the start depot is $\CustCount + 1$ and the end depot is $0$.  To initialize $\Last_\TimeStart$ in Step~\ref{alg:tsptw-d:init-t}, we run Algorithm~\ref{alg:basic} on $\Network^{-1}$ with $\Time_0 = 0$ and $\UpperBound = \Horizon - \Time^*$.  If a route $\Route^{-1}$ in $\Network^{-1}$ is found, then $\Last_\TimeStart = \Horizon - \ArrivalTime(\Route^{-1}, 0)$; otherwise, $\Last_\TimeStart = \Time^*$.  Finally, we use a local search heuristic for Steps~\ref{alg:tsptw-d:local search 1}~and~\ref{alg:tsptw-d:local search 2},  which applies the swap, $2$-opt, and shift operators with a first-improvement strategy.  Our implementation is not particularly efficient, as it spends $O(n^2)$ time on each operator.

\section{Computational Results}
\label{sec:basic:computational results}

We implemented Algorithms~\ref{alg:bfs}--\ref{alg:tsptw-d} in \CPP{} to evaluate their performance on the benchmark instances from Section~\ref{sec:problem statement:benchmarks}. The code is freely available at \url{https://github.com/fsoulignac/tsptw-m}.  We executed our experiments on a single thread of a laptop equipped with an AMD Ryzen 7 3700U CPU@$2.3$GHz and $6$GB of physical RAM.  To limit the impact of paging on the running time, we restricted the available RAM to $5$GB.

\subsection{Results for the \TSPTWM{}}

For the \TSPTWM{} we set a time limit of $3$ minutes per instance. Although this time limit is quite restrictive for an exact method, it is sufficient for our purposes and helps prevent unnecessary computation time before encountering a memory limit exception.  Furthermore, this time limit is representative of a typical use case, where an enumerative algorithm is applied for a short duration after computing a tight lower bound \citep[see e.g.][]{BaldacciMingozziRoberti2012,TilkIrnich2017,Lera-RomeroMirandaBrontSoulignac2022}.

The results for the classical benchmarks and \DaSilvaUrrutia{} are summarized in Table~\ref{tab:classical basic}.  We compare Algorithm~\ref{alg:basic} against the methods by \cite{Lera-RomeroMirandaBrontSoulignac2022} (\LeraRomero), \cite{RudichCappartRousseau2023} (\Rudich), and \cite{FontaineDibangoyeSolnon2023} (\Fontaine), which represent the state of the art for the \TSPTWM{} within exact methods. \LeraRomero{} and \Fontaine{} were originally proposed for time-dependent versions of the TSPTW, and thus \cite{Lera-RomeroMirandaBrontSoulignac2022} and \cite{FontaineDibangoyeSolnon2023} did not report the results for the \TSPTWM{}.  Similarly, \cite{RudichCappartRousseau2023} did not test \Rudich{} on the \DaSilvaUrrutia{} benchmark.  These missing results reported in Table~\ref{tab:classical basic} are taken from \cite{Fontaine2024}.  We note that the results in \cite{Fontaine2024} differ from those reported by \cite{RudichCappartRousseau2023} because \cite{Fontaine2024} executed all the algorithms on the same machine and with the same configuration.  Our computer has a slightly higher performance for single thread execution according to \url{www.cpubenchmark.net}, and, unlike \cite{Fontaine2024}, we did not disable turbo mode (we kept the ondemand governor within Ubuntu). Additionally, the time limit considered by \cite{Fontaine2024} was one hour per instance.   These discrepancies do not affect our main conclusions, and we thus disregard them.

In Table~\ref{tab:classical basic}, column \# counts the number of instances in each benchmark set, columns $s$ indicate the number of solved instances, and columns $t_s$ report the average time for the solved instances, rounded to the nearest second. For Algorithm~\ref{alg:basic}, we also include the maximum running time among the solved instances in column $m_s$.  Additionally, we divide the results for \Ascheuer{} into \sAscheuer{} and \lAscheuer{} to separately report the results for instances with fewer than $50$ customers and those with $50$ or more customers, respectively.  Table~\ref{tab:classical basic} reports the results for \Rudich{} with the unseeded version outside parentheses and the seeded version inside parentheses.  The difference between the two is that the latter is initialized with a known solution. In this regard, Algorithm~\ref{alg:basic} should be considered unseeded.  Similarly, Table~\ref{tab:classical basic} reports the results for \Fontaine{} for two versions of the heuristic bound guiding the informed search: FEA (feasiblity) bound and MSA (minimum spanning arborescence) bound.  The results for FEA are reported outside parentheses, and those for MSA are reported inside parentheses.

\begin{table}
    \centering
    \small
    \begin{tabular}{lc ccc ccc ccc ccc}
        \toprule
         &    & \multicolumn{2}{c}{\LeraRomero} && \multicolumn{2}{c}{\Rudich} && \multicolumn{2}{c}{\Fontaine} && \multicolumn{3}{c}{Algorithm~\ref{alg:basic}} \\\cmidrule{3-4}\cmidrule{6-7}\cmidrule{9-10}\cmidrule{12-14}
                        & \#  & $s^*$ & $t_s^*$ && $s$      & $t_s$     && $s^*$    & $t_s^*$ && $s$ & $t_s$ & $m_s$ \\\midrule
        \Ascheuer       & 50  & 48    & 21      && 48(48)   & 173(109)  && 50(49)   & 18(2)   && 50  & 13    & 152   \\
        \sAscheuer      & 32  &       &         &&          &           &&          &         && 32  & 21    & 152   \\
        \lAscheuer      & 18  &       &         &&          &           &&          &         && 18  &  0    &   0   \\
        \DaSilvaUrrutia & 125 & 125   & 485     && 112*     & 615*      && 125(125) & 13(13)  && 125 & 2     & 7     \\
        \Dumas          & 135 & 135   & 39      && 133(135) & 154(111)  && 135(135) & 0(1)    && 135 & 0     & 1     \\
        \Gendreau       & 130 & 91    & 524     && 122(125) & 274(142)  && 117(110) & 111(88) && 130 & 0     & 1     \\
        \Langevin       & 70  & 70    & 0       && 70(70)   & 1(1)      && 70(70)   & 0(0)    && 70  & 0     & 0     \\
        \OhlmannThomas  & 25  & 0     & ---     && 14(22)   & 1888(924) && 20(20)   & 50(9)   && 25  & 1     & 6     \\
        \Pesant         & 27  & 22    & 203     && 26(27)   & 175(84)   && 25(27)   & 152(66) && 25  & 16    & 164   \\
        \PotvinBengio   & 30  & 25    & 160     && 28(28)   & 185(66)   && 27(29)   & 14(83)  && 25  & 7     & 80    \\\midrule
        Total & 592 & 516 &&& 553(567) &&& 569(565) &&& 585\\\bottomrule
    \end{tabular}
    \caption{Results of Algorithm~\ref{alg:basic} on the classical benchmarks.  Entries marked with * were reported by \cite{Fontaine2024}.}\label{tab:classical basic}
\end{table}

A quick look at Table~\ref{tab:classical basic} is enough to conclude that Algorithm~\ref{alg:basic} outperforms the state-of-the-art methods on the larger instances, solving the four instances that remained unsolved.   Although this is strictly true, knowing the inner workings of Algorithm~\ref{alg:basic}, we would never prefer it as a generic exact solver for the \TSPTWM{}.  A deeper look at Table~\ref{tab:classical basic} raises an important concern: Algorithm~\ref{alg:basic} is unable to solve seven instances with 45 or fewer customers (and our unreported experiments show that memory is exhausted when the time limit is high enough).  Furthermore, Algorithm~\ref{alg:basic} performs better on the larger instances in \Ascheuer{} than on the smaller ones.  As strange as it may seem, \cite{RifkiSolnon2025} already noted that the ratio between the size of the time windows and the horizon could be more important than the number of customers when evaluating the efficiency of dynamic programming solvers for the \TSPTWM{}.  Finally, we observe that Algorithm~\ref{alg:basic} is particularly fast on \Langevin, \Dumas, \Gendreau, \OhlmannThomas, and \DaSilvaUrrutia, all of which follow the same creation procedure commonly applied in many machine learning methods to create ``hard'' training datasets (Section~\ref{sec:problem statement:benchmarks}).

What we believe is happening is that Algorithm~\ref{alg:basic} takes advantage of some inherent bias in the benchmark instances resulting from the tight time windows, which yield a small search space of feasible solutions that have a well defined structure.  \Fontaine{} is another informed search that seems to have an advantage because of the structure of the instances, as depicted by the \OhlmannThomas{} benchmark: the $20$ instances it solves are completed in just a few seconds, while \LeraRomero{} is unable to solve any of them within an hour!  The difference is that \LeraRomero{} was designed to solve as many as possible of the small but challenging instances.  As a result, it spends a large amount of time computing penalties on a relaxed problem before running its enumeration algorithm, and it uses a pure best-first search heuristic to guide its informed search, avoiding the extension of dominated labels.  Finally, \Rudich{} is a generic algorithm based on multivalued decision diagrams that iteratively traverses enumeration trees of different widths to obtain lower and upper bounds to ``peel'' the tree. Being able to find feasible solutions quickly improves the peeling process.

To present the contrasting results, we report the performance of the algorithms for the \Rifki{} benchmark in Table~\ref{tab:rifki basic}. \citep[The results for \LeraRomero{}, \Rudich{}, and \Fontaine{} were taken from][]{Fontaine2024}. Again, Algorithm~\ref{alg:basic} is among the fastest for tight time windows ($\TWTightness \in \{50,100\}$), but it is clearly the worst for the looser time windows ($\TWTightness \in \{0,25\}$), even when compared against the generic method \Rudich{}. (We remark that increasing the time limit does not help in these cases, as the memory limit is always reached.) The clear winner in robustness is \LeraRomero{}, which was specifically designed for harder instances in terms of time windows.

\begin{table}
    \centering
    \small
    \begin{tabular}{cc ccc ccc ccc ccc}
        \toprule
        $\TWTightness$ & $\CustCount$ & \multicolumn{2}{c}{\LeraRomero} && \multicolumn{2}{c}{\Rudich} && \multicolumn{2}{c}{\Fontaine} && \multicolumn{3}{c}{Algorithm~\ref{alg:basic}} \\\cmidrule{3-4}\cmidrule{6-7}\cmidrule{9-10}\cmidrule{12-14}
              &         & $s$ & $t_s$ && $s$ & $t_s$ && $s$ & $t_s$ && $s$ & $t_s$ & $m_s$ \\\midrule
        $100$ & $20$    & 60  & 0     && 60  & 0     && 60  & 0     && 60  & 0     & 0     \\
              & $30$    & 60  & 0     && 60  & 1     && 60  & 0     && 60  & 0     & 0     \\
              & $40$    & 60  & 0     && 60  & 1     && 60  & 0     && 60  & 0     & 0     \\\midrule
        $50$  & $20$    & 60  & 5     && 60  & 9     && 60  & 0     && 60  & 0     & 2     \\
              & $30$    & 60  & 129   && 58  & 105   && 60  & 1     && 59  & 5     & 125   \\
              & $40$    & 60  & 888   && 14  & 350   && 60  & 61    && 58  & 18    & 117   \\\midrule
        $25$  & $20$    & 60  & 16    && 60  & 17    && 60  & 0     && 60  & 10    & 60    \\
              & $30$    & 60  & 643   && 27  & 361   && 60  & 36    &&  0  & ---   & ---   \\
              & $40$    & 58  & 2318  && 0   & ---   && 47  & 1221  &&  0  & ---   & ---   \\\midrule
        $0$   & $20$    & 60  & 59    && 60  & 25    && 60  & 0     && 60  & 35    & 177   \\
              & $30$    & 60  & 1376  && 10  & 434   && 60  & 177   &&  0  & ---   & ---   \\
              & $40$    & 34  & 2837  && 0   & ---   && 5   & 2011  &&  0  & ---   & ---   \\\midrule
        Total &  & 692 &&& 469 &&& 652 &&& 506\\\bottomrule
    \end{tabular}
    \caption{Results of Algorithm~\ref{alg:basic} on the \Rifki{} benchmark.}\label{tab:rifki basic}
\end{table}

In our opinion, the main conclusion from the experiments in this section is that the classical benchmarks are no longer adequate when used as the sole tool for evaluating the \TSPTWM{}.  If not complemented by other benchmarks, any method that incorporates a variant of Algorithm~\ref{alg:basic} in its preprocessing step would likely produce results that may appear outstanding to an external reviewer.

\subsection{Results for the \TSPTWD{}}
\label{sec:tsptw-d:computational results}

For the \TSPTWD{}, we set a time limit of $30$ minutes per instance.   Table~\ref{tab:tsptw-d} summarizes the results for the benchmarks \Ascheuer{}, \DaSilvaUrrutia{}, \Gendreau{}, \OhlmannThomas{}, and \DOhlmannThomas.  It includes the results of Algorithm~\ref{alg:tsptw-d} applied to the transport network $\Network$ (columns labeled Algorithm~\ref{alg:tsptw-d}) and its reverse network $\Network^{-1}$ (columns labeled Algorithm~\ref{alg:tsptw-d}$^{-1}$), as the method is not symmetric, and the running times differ considerably between the two.  For comparison, Table~\ref{tab:tsptw-d} also reports results from \cite{TilkIrnich2017} and \cite{Lera-RomeroMirandaBrontSoulignac2022}.

We do not consider the benchmarks \PotvinBengio{} and \Pesant{} because their travel times are rounded to $10^{-4}$, and Algorithm~\ref{alg:tsptw-d} cannot handle the large number of \TSPTWM{} instances in the range $[10^{4}\Horizon]$.  This also applies to the four unsolved instances in \Ascheuer{} (rbg021.$x$ for $x \in \Range{6,9}$), which were derived from rbg021 by increasing both the horizon and the deadline times in the time windows.  Additionally, we exclude the benchmark \Langevin{} because its instances are trivial and were not tested by \cite{TilkIrnich2017} or \cite{Lera-RomeroMirandaBrontSoulignac2022}. We also note that \cite{TilkIrnich2017,Lera-RomeroMirandaBrontSoulignac2022} did not evaluate their algorithms on the \Gendreau{} instances with $\CustCount \geq 80$ and $\TWWidth \in \{80,100\}$.  It is worth mentioning that \cite{Lera-RomeroMirandaBrontSoulignac2022} solved all the instances in \PotvinBengio{}, while \cite{TilkIrnich2017} solved all but three. However, both methods were tested with a time limit of 3 hours per instance.

\begin{table}
    \centering
    \small
    \begin{tabular}{lr rrr rrr rrrr rrr}
        \toprule
         &    & \multicolumn{2}{c}{\Tilk} && \multicolumn{2}{c}{\LeraRomero} && \multicolumn{3}{c}{Algorithm~\ref{alg:tsptw-d}} && \multicolumn{3}{c}{Algorithm~\ref{alg:tsptw-d}$^{-1}$}
         \\\cmidrule{3-4}\cmidrule{6-7}\cmidrule{9-11}\cmidrule{13-15}
                        & \#  & $s$     & $t_s$ && $s$     & $t_s$ && $s$ & $t_s$ & $m_s$ && $s$ & $t_s$ & $m_s$ \\\midrule
        \Ascheuer       &  50 & 50      &   55  && 47      &   8   &&  46 &   9   & 295   &&  45 &   3   &  60   \\
        \DaSilvaUrrutia & 125 & ---     &  ---  && ---     & ---   && 125 &   2   &   7   && 125 &   2   &   8   \\
        \Dumas          & 135 & ---     &  ---  && ---     & ---   && 135 &   0   &   1   && 135 &   0   &   1   \\
        \Gendreau       & 130 &  97/115 &  633  && 114/115 &  667  && 123 &  30   & 1279  && 129 &   3   & 173   \\
        \OhlmannThomas  &  25 & ---     &  ---  && ---     & ---   &&  25 &  15   &  216  &&  25 &   3   &  27   \\
        \DOhlmannThomas &  25 &  1      &  731  &&  12     & 5159  &&  20 &  10   &   59  &&  25 &  92   & 1093  \\\bottomrule
    \end{tabular}
    \caption{Results of Algorithm~\ref{alg:tsptw-d} on the classical benchmarks.}\label{tab:tsptw-d}
\end{table}

Table~\ref{tab:tsptw-d} reinforces the conclusions from Section~\ref{sec:basic:computational results}:  Algorithm~\ref{alg:tsptw-d} (on either $\Network$ or $\Network^{-1}$) solves all but one of the classical instances with $50$ or more customers, including the 14 instances that were not solved by \cite{TilkIrnich2017} or \cite{Lera-RomeroMirandaBrontSoulignac2022}.  Furthermore, Algorithm~\ref{alg:tsptw-d} outperforms both methods on the larger instances with higher values of $\CustCount$.

By design, one might expect Algorithm~\ref{alg:tsptw-d} to be much less efficient than Algorithm~\ref{alg:basic}, as it solves $\Theta(\Horizon)$ instances of the \TSPTWM{} in the worst case.  However, as implied by Table~\ref{tab:tsptw-d}, only a small number of these instances are actually passed to Algorithm~\ref{alg:basic} for solving in the benchmark instances. Most feasible departure times $\TimeStart$ in Loop~\ref{alg:tsptw-d:loop start}--\ref{alg:tsptw-d:loop end} are either solved by the local search heuristic or ignored because of the best-known solution $\Route^*$.  Moreover, $\Last_\TimeStart$ is typically small in the largest instances, which helps reduce the number of instances solved by Algorithm~\ref{alg:basic}.  In short, solving the \TSPTWD{} on these instances is not substantially harder than solving the \TSPTWM{}.

The main conclusion is that the evaluation of the \TSPTWD{} should be complemented with more difficult instances, especially those with wider time windows, to better assess the performance of algorithms in challenging scenarios.

\section{Conclusions}
\label{sec:conclusions}

We presented a simple informed search method for the \TSPTWM{}, which excels in solving all classical benchmark instances with 50 or more customers in under 10 seconds each. However, it struggles to solve even the smallest instances when their time windows are sufficiently loose. By running the algorithm as a preprocessing step for a limited amount of time, any method can solve these instances, meaning it is generally advisable to complement them with harder instances featuring looser time windows, rather than using them alone for benchmarking.

An important question that arises from this work is how valid and accurate are the conclusions drawn from methods that have been tested exclusively on these classical benchmark instances? Are these methods truly efficient in a general setting, or are they, perhaps unconsciously, exploiting some inherent bias in these benchmarks? Since these instances appear easier for simple enumerative methods, it is crucial to examine how generalizable the reported results are. Exploring this issue further could help ensure that conclusions drawn from these benchmarks are not only valid for the specific instances tested but also robust across a wider range of real-world problem scenarios.

Our method performs particularly well on the benchmarks \Langevin, \Dumas, \Gendreau, \OhlmannThomas, and \DaSilvaUrrutia, all of which share the common strategy of defining a time window $[x - \TWWidth, x + \TWWidth]$ for a vertex visited at time $x$ in a given route. Here, $\TWWidth$ is relatively small when compared to $\CustCount$ and $\Horizon$. Due to the simplicity of this approach, it is commonly used to generate training datasets for many machine learning methods, which can lead to overfitting and biased conclusions. As noted by \cite{RifkiSolnon2025}, a simple solution to make these instances more challenging for an informed search algorithm like ours is to extend the time windows. The varying levels of difficulty among these generated instances provide a better foundation for drawing deeper conclusions.

Given the exceptional performance of our method on the larger benchmark instances, it is not surprising that a trivial procedure of repeatedly running the solver across the entire horizon is sufficient to solve the \TSPTWD{} on most of the instances. The issue is that many of these instances are easy even for the \TSPTWD{}. Therefore, these instances are not sufficient when evaluating methods for the \TSPTWD{}. What remains to be explored is whether a similar informed search can be applied to solve the \TSPTWTT{} on these instances. Nonetheless, we believe it would be wise to evaluate and train future methods on instances with varying levels of time window tightness, even for the \TSPTWTT{}.

\bibliographystyle{abbrvnat}
\bibliography{simple-tsptw-solver}

@Article{TilkIrnich2017,
  author   = {Tilk, Christian and Irnich, Stefan},
  title    = {Dynamic Programming for the Minimum Tour Duration Problem},
  journal  = {Transp. Sci.},
  year     = {2017},
  volume   = {51},
  number   = {2},
  pages    = {549-565},
  doi      = {10.1287/trsc.2015.0626},
  fjournal = {Transportation Science},
}

@Article{Lopez-IbanezBlumOhlmannThomas2013,
  author   = {Manuel López-Ibáñez and Christian Blum and Jeffrey W. Ohlmann and Barrett W. Thomas},
  title    = {The travelling salesman problem with time windows: Adapting algorithms from travel-time to makespan optimization},
  journal  = {Appl. Soft Comput.},
  year     = {2013},
  volume   = {13},
  number   = {9},
  pages    = {3806-3815},
  doi      = {10.1016/j.asoc.2013.05.009},
  fjournal = {Applied Soft Computing},
}

@Article{ChristofidesMingozziToth1981,
  author   = {Christofides, Nicos and Mingozzi, A. and Toth, P.},
  title    = {State-space relaxation procedures for the computation of bounds to routing problems},
  journal  = {Netw.},
  year     = {1981},
  volume   = {11},
  number   = {2},
  pages    = {145-164},
  doi      = {10.1002/net.3230110207},
  fjournal = {Networks},
}

@Article{Baker1983,
  author   = {Baker, Edward K.},
  title    = {Technical Note—An Exact Algorithm for the Time-Constrained Traveling Salesman Problem},
  journal  = {Oper. Res.},
  year     = {1983},
  volume   = {31},
  number   = {5},
  pages    = {938-945},
  doi      = {10.1287/opre.31.5.938},
  fjournal = {Operations Research},
}

@Article{DumasDesrosiersGelinasSolomon1995,
  author   = {Dumas, Yvan and Desrosiers, Jacques and Gelinas, Eric and Solomon, Marius M.},
  title    = {An Optimal Algorithm for the Traveling Salesman Problem with Time Windows},
  journal  = {Oper. Res.},
  year     = {1995},
  volume   = {43},
  number   = {2},
  pages    = {367-371},
  doi      = {10.1287/opre.43.2.367},
  fjournal = {Operations Research},
}

@Article{LangevinDesrochersDesrosiersGelinasSoumis1993,
  author   = {Langevin, André and Desrochers, Martin and Desrosiers, Jacques and Gélinas, Sylvie and Soumis, Fraňlois},
  title    = {A two-commodity flow formulation for the traveling salesman and the makespan problems with time windows},
  journal  = {Netw.},
  year     = {1993},
  volume   = {23},
  number   = {7},
  pages    = {631-640},
  doi      = {10.1002/net.3230230706},
  fjournal = {Networks},
}

@Article{AscheuerFischettiGroetschel2001,
  author   = {Ascheuer, Norbert and Fischetti, Matteo and Grötschel, Martin},
  title    = {Solving the Asymmetric Travelling Salesman Problem with time windows by branch-and-cut},
  journal  = {Math. Program.},
  year     = {2001},
  volume   = {90},
  pages    = {475--506},
  doi      = {10.1007/PL00011432},
  fjournal = {Mathematical Programming},
  issue    = {3},
}

@Article{BalasSimonetti2001,
  author   = {Balas, Egon and Simonetti, Neil},
  title    = {Linear Time Dynamic-Programming Algorithms for New Classes of Restricted {TSPs}: A Computational Study},
  journal  = {INFORMS J. Comput.},
  year     = {2001},
  volume   = {13},
  number   = {1},
  pages    = {56-75},
  doi      = {10.1287/ijoc.13.1.56.9748},
  fjournal = {INFORMS Journal on Computing},
}

@Article{FocacciLodiMilano2002,
  author   = {Focacci, Filippo and Lodi, Andrea and Milano, Michela},
  title    = {A Hybrid Exact Algorithm for the {TSPTW}},
  journal  = {INFORMS J. Comput.},
  year     = {2002},
  volume   = {14},
  number   = {4},
  pages    = {403-417},
  doi      = {10.1287/ijoc.14.4.403.2827},
  fjournal = {INFORMS Journal on Computing},
}

@Article{CarltonBarnes1996,
  author    = {William B. Carlton and J. Wesley Barnes},
  title     = {Solving the Traveling-Salesman Problem with Time Windows Using Tabu Search},
  journal   = {IIE Trans.},
  year      = {1996},
  volume    = {28},
  number    = {8},
  pages     = {617--629},
  doi       = {10.1080/15458830.1996.11770707},
  fjournal  = {IIE Transactions},
  publisher = {Taylor \& Francis},
}

@Article{GendreauHertzLaporteStan1998,
  author   = {Gendreau, Michel and Hertz, Alain and Laporte, Gilbert and Stan, Mihnea},
  title    = {A Generalized Insertion Heuristic for the Traveling Salesman Problem with Time Windows},
  journal  = {Oper. Res.},
  year     = {1998},
  volume   = {46},
  number   = {3},
  pages    = {330-335},
  doi      = {10.1287/opre.46.3.330},
  fjournal = {Operations Research},
}

@Article{Calvo2000,
  author   = {Calvo, Roberto Wolfler},
  title    = {A New Heuristic for the Traveling Salesman Problem with Time Windows},
  journal  = {Transp. Sci.},
  year     = {2000},
  volume   = {34},
  number   = {1},
  pages    = {113-124},
  doi      = {10.1287/trsc.34.1.113.12284},
  fjournal = {Transportation Science},
}

@Article{daSilvaUrrutia2010,
  author   = {Rodrigo Ferreira {da Silva} and Sebastián Urrutia},
  title    = {A General {VNS} heuristic for the traveling salesman problem with time windows},
  journal  = {Discrete Optim.},
  year     = {2010},
  volume   = {7},
  number   = {4},
  pages    = {203-211},
  doi      = {10.1016/j.disopt.2010.04.002},
  fjournal = {Discrete Optimization},
}

@Article{FavarettoMorettiPellegrini2006,
  author    = {Daniela Favaretto and Elena Moretti and Paola Pellegrini},
  title     = {An ant colony system approach for variants of the traveling salesman problem with time windows},
  journal   = {J. Inf. Optim. Sci.},
  year      = {2006},
  volume    = {27},
  number    = {1},
  pages     = {35--54},
  doi       = {10.1080/02522667.2006.10699677},
  publisher = {Taylor \& Francis},
}

@Article{KaraKocAltiparmakDengiz2013,
  author    = {Imdat Kara and Ozge Nimet Koc and Fulya Altıparmak and Berna Dengiz},
  title     = {New integer linear programming formulation for the traveling salesman problem with time windows: minimizing tour duration with waiting times},
  journal   = {Optim.},
  year      = {2013},
  volume    = {62},
  number    = {10},
  pages     = {1309--1319},
  doi       = {10.1080/02331934.2013.824445},
  fjournal  = {Optimization},
  publisher = {Taylor \& Francis},
}

@Article{Savelsbergh1992,
  author  = {Savelsbergh, Martin W. P.},
  title   = {The Vehicle Routing Problem with Time Windows: Minimizing Route Duration},
  journal = {ORSA J. Comput.},
  year    = {1992},
  volume  = {4},
  number  = {2},
  pages   = {146-154},
  doi     = {10.1287/ijoc.4.2.146},
}

@InProceedings{GillardCoppeSchausCire2021,
  author    = {Gillard, Xavier and Copp{\'e}, Vianney and Schaus, Pierre and Cire, Andr{\'e} Augusto},
  title     = {Improving the Filtering of Branch-and-Bound {MDD} Solver},
  booktitle = {Integration of Constraint Programming, Artificial Intelligence, and Operations Research},
  year      = {2021},
  editor    = {Stuckey, Peter J.},
  pages     = {231--247},
  publisher = {Springer International Publishing},
  doi       = {10.1007/978-3-030-78230-6_15},
}

@Article{Pralet2023,
  author   = {Cédric Pralet},
  title    = {Iterated Maximum Large Neighborhood Search for the Traveling Salesman Problem with Time Windows and its Time-dependent Version},
  journal  = {Comput. Oper. Res.},
  year     = {2023},
  volume   = {150},
  pages    = {106078},
  doi      = {10.1016/j.cor.2022.106078},
  fjournal = {Computers \& Operations Research},
}

@TechReport{AmgharCordeauGendron2019,
  author      = {Khalid Amghar and Jean-François Cordeau and Bernard Gendron},
  title       = {A General Variable Neighborhood Search Heuristic for the Traveling Salesman Problem with Time Windows under Completion Time Minimization},
  institution = {CIRRELT},
  year        = {2019},
}

@Article{RudichCappartRousseau2023,
  author   = {Isaac Rudich and Quentin Cappart and Louis{-}Martin Rousseau},
  title    = {Improved Peel-and-Bound: Methods for Generating Dual Bounds with Multivalued Decision Diagrams},
  journal  = {J. Artif. Intell. Res.},
  year     = {2023},
  volume   = {77},
  pages    = {1489--1538},
  doi      = {10.1613/JAIR.1.14607},
  fjournal = {Journal of Artificial Intelligence Research},
}

@InProceedings{BolandHewittVuSavelsbergh2017,
  author    = {Boland, Natashia and Hewitt, Mike and Vu, Duc Minh and Savelsbergh, Martin},
  title     = {Solving the Traveling Salesman Problem with Time Windows Through Dynamically Generated Time-Expanded Networks},
  booktitle = {Integration of AI and OR Techniques in Constraint Programming},
  year      = {2017},
  editor    = {Salvagnin, Domenico and Lombardi, Michele},
  pages     = {254--262},
  publisher = {Springer International Publishing},
  doi       = {10.1007/978-3-319-59776-8_21},
}

@Article{MladenovicTodosijevicUrosevic2013,
  author   = {Mladenović, Nenad and Todosijević, Raca and Urošević, Dragan},
  title    = {An efficient General Variable Neighborhood Search for large Travelling Salesman Problem with Time Windows},
  journal  = {Yugosl. J. Oper. Res.},
  year     = {2013},
  volume   = {23},
  pages    = {19--30},
  doi      = {10.2298/YJOR120530015M},
  fjournal = {Yugoslav Journal of Operations Research},
  issue    = {1},
}

@Article{BaldacciMingozziRoberti2012,
  author   = {Baldacci, Roberto and Mingozzi, Aristide and Roberti, Roberto},
  title    = {New State-Space Relaxations for Solving the Traveling Salesman Problem with Time Windows},
  journal  = {INFORMS J. Comput.},
  year     = {2012},
  volume   = {24},
  number   = {3},
  pages    = {356-371},
  doi      = {10.1287/ijoc.1110.0456},
  fjournal = {INFORMS Journal on Computing},
}

@Article{YeBartoliniSchneider2024,
  author   = {Mengdie Ye and Enrico Bartolini and Michael Schneider},
  title    = {A general variable neighborhood search for the traveling salesman problem with time windows under various objectives},
  journal  = {Discrete Appl. Math.},
  year     = {2024},
  volume   = {346},
  pages    = {95-114},
  doi      = {10.1016/j.dam.2023.12.006},
  fjournal = {Discrete Applied Mathematics},
}

@PhdThesis{Fontaine2024,
  author = {Fontaine, Romain},
  title  = {Exact and anytime heuristic search for the Time Dependent Traveling Salesman Problem with Time Windows},
  school = {{INSA de Lyon}},
  year   = {2024},
  url = {https://hal.science/tel-04697323},
  number = {2024ISAL0067},
  month = Jul,
  type = {Theses},
  HAL_ID = {tel-04697323},
  HAL_VERSION = {v3},
}

@Article{Lera-RomeroMirandaBrontSoulignac2022,
  author   = {Lera-Romero, Gonzalo and Miranda Bront, Juan Jos\'{e} and Soulignac, Francisco J.},
  title    = {Dynamic Programming for the Time-Dependent Traveling Salesman Problem with Time Windows},
  journal  = {INFORMS J. Comput.},
  year     = {2022},
  volume   = {34},
  number   = {6},
  pages    = {3292-3308},
  doi      = {10.1287/ijoc.2022.1236},
  fjournal = {INFORMS Journal on Computing},
}

@PhdThesis{Ascheuer1996,
  author = {Norbert Ascheuer},
  title  = {Hamiltonian path problems in the on-line optimization of flexible manufacturing systems},
  school = {Zuse Institute Berlin},
  year   = {1996},
}

@Article{OhlmannThomas2007,
  author   = {Ohlmann, Jeffrey W. and Thomas, Barrett W.},
  title    = {A Compressed-Annealing Heuristic for the Traveling Salesman Problem with Time Windows},
  journal  = {INFORMS J. Comput.},
  year     = {2007},
  volume   = {19},
  number   = {1},
  pages    = {80-90},
  fjournal = {INFORMS Journal on Computing},
}

@Article{PotvinBengio1996,
  author   = {Potvin, Jean-Yves and Bengio, Samy},
  title    = {The Vehicle Routing Problem with Time Windows Part {II}: Genetic Search},
  journal  = {INFORMS J. Comput.},
  year     = {1996},
  volume   = {8},
  number   = {2},
  pages    = {165-172},
  doi      = {10.1287/ijoc.8.2.165},
  fjournal = {INFORMS Journal on Computing},
}

@Article{PesantGendreauPotvinRousseau1998,
  author   = {Pesant, Gilles and Gendreau, Michel and Potvin, Jean-Yves and Rousseau, Jean-Marc},
  title    = {An Exact Constraint Logic Programming Algorithm for the Traveling Salesman Problem with Time Windows},
  journal  = {Transp. Sci},
  year     = {1998},
  volume   = {32},
  number   = {1},
  pages    = {12-29},
  doi      = {10.1287/trsc.32.1.12},
  fjournal = {Transportation Science},
}

@Article{RifkiChiabautSolnon2020,
  author   = {Omar Rifki and Nicolas Chiabaut and Christine Solnon},
  title    = {On the impact of spatio-temporal granularity of traffic conditions on the quality of pickup and delivery optimal tours},
  journal  = {Transp. Res. Part E},
  year     = {2020},
  volume   = {142},
  pages    = {102085},
  doi      = {10.1016/j.tre.2020.102085},
  fjournal = {Transportation Research Part E: Logistics and Transportation Review},
}

@Article{AriglianoGhianiGriecoGuerrieroPlana2019,
  author   = {Anna Arigliano and Gianpaolo Ghiani and Antonio Grieco and Emanuela Guerriero and Isaac Plana},
  title    = {Time-dependent asymmetric traveling salesman problem with time windows: Properties and an exact algorithm},
  journal  = {Discret. Appl. Math.},
  year     = {2019},
  volume   = {261},
  pages    = {28-39},
  doi      = {10.1016/j.dam.2018.09.017},
  fjournal = {Discrete Applied Mathematics},
}

@Article{FontaineDibangoyeSolnon2023,
  author   = {Romain Fontaine and Jilles Dibangoye and Christine Solnon},
  title    = {Exact and anytime approach for solving the time dependent traveling salesman problem with time windows},
  journal  = {Eur. J. Oper. Res.},
  year     = {2023},
  volume   = {311},
  number   = {3},
  pages    = {833-844},
  doi      = {10.1016/j.ejor.2023.06.001},
  fjournal = {European Journal of Operational Research},
}

@Article{Savelsbergh1985,
  author   = {Savelsbergh, M. W. P.},
  title    = {Local search in routing problems with time windows},
  journal  = {Ann. Oper. Res.},
  year     = {1985},
  volume   = {4},
  pages    = {285--305},
  doi      = {10.1007/BF02022044},
  fjournal = {Annals of Operations Research},
  ussue    = {1},
}

@Article{MingozziBiancoRicciardelli1997,
  author   = {Mingozzi, Aristide and Bianco, Lucio and Ricciardelli, Salvatore},
  title    = {Dynamic Programming Strategies for the Traveling Salesman Problem with Time Window and Precedence Constraints},
  journal  = {Oper. Res.},
  year     = {1997},
  volume   = {45},
  number   = {3},
  pages    = {365-377},
  doi      = {10.1287/opre.45.3.365},
  fjournal = {Operations Research},
}

@Article{Lopez-IbanezBlum2010,
  author   = {Manuel López-Ibáñez and Christian Blum},
  title    = {Beam-{ACO} for the travelling salesman problem with time windows},
  journal  = {Comput. Oper. Res.},
  year     = {2010},
  volume   = {37},
  number   = {9},
  pages    = {1570-1583},
  doi      = {10.1016/j.cor.2009.11.015},
  fjournal = {Computers \& Operations Research},
}

@Article{KuroiwaBeck2023,
  author   = {Kuroiwa, Ryo and Beck, J. Christopher},
  title    = {Solving Domain-Independent Dynamic Programming Problems with Anytime Heuristic Search},
  journal  = {Proc. Int. Conf. Autom. Plan. Sched.},
  year     = {2023},
  volume   = {33},
  number   = {1},
  pages    = {245-253},
  doi      = {10.1609/icaps.v33i1.27201},
  fjournal = {Proceedings of the International Conference on Automated Planning and Scheduling},
}

@Article{CoppeGillardSchaus2024,
  author   = {Copp\'{e}, Vianney and Gillard, Xavier and Schaus, Pierre},
  title    = {Decision Diagram-Based Branch-and-Bound with Caching for Dominance and Suboptimality Detection},
  journal  = {INFORMS J. Comput.},
  year     = {2024},
  volume   = {36},
  number   = {6},
  pages    = {1522-1542},
  doi      = {10.1287/ijoc.2022.0340},
  fjournal = {INFORMS Journal on Computing},
}

@Article{RifkiSolnon2025,
  author   = {Omar Rifki and Christine Solnon},
  title    = {On the Phase Transition of the Euclidean Travelling Salesman Problem with Time Windows},
  journal  = {J. Artif. Intell. Res.},
  year     = {2025},
  volume   = {82},
  pages    = {2167--2188},
  doi      = {10.1613/jair.1.18334},
  fjournal = {Journal of Artificial Intelligence Research},
}

@Misc{li2025lmasklearnsolveconstrained,
  author        = {Tianyou Li and Haijun Zou and Jiayuan Wu and Zaiwen Wen},
  title         = {{LMask}: Learn to Solve Constrained Routing Problems with Lazy Masking},
  year          = {2025},
  archiveprefix = {arXiv},
  eprint        = {2505.17938},
  primaryclass  = {math.OC},
  url           = {https://arxiv.org/abs/2505.17938},
}

@InProceedings{ChenGongChenLiuWangYuZhang2025,
  author    = {Chen, Jingxiao and Gong, Ziqin and Chen, Lvda and Liu, Minghuan and Wang, Jun and Yu, Yong and Zhang, Weinan},
  title     = {Looking Ahead to Avoid Being Late: Solving Hard-Constrained Traveling Salesman Problem},
  booktitle = {Proceedings of the 2024 6th International Conference on Distributed Artificial Intelligences},
  year      = {2025},
  series    = {DAI '24},
  pages     = {1–12},
  publisher = {Association for Computing Machinery},
  doi       = {10.1145/3719545.3759088},
}

@InProceedings{KoolHoofGromichoWelling2022,
  author    = {Kool, Wouter and van Hoof, Herke and Gromicho, Joaquim and Welling, Max},
  title     = {Deep Policy Dynamic Programming for Vehicle Routing Problems},
  booktitle = {Integration of Constraint Programming, Artificial Intelligence, and Operations Research: 19th International Conference, CPAIOR 2022, Los Angeles, CA, USA, June 20-23, 2022, Proceedings},
  year      = {2022},
  pages     = {190–213},
  address   = {Berlin, Heidelberg},
  publisher = {Springer-Verlag},
  doi       = {10.1007/978-3-031-08011-1_14},
}

@InProceedings{BiMaZhouSongCaoWuZhang2024,
  author    = {Bi, Jieyi and Ma, Yining and Zhou, Jianan and Song, Wen and Cao, Zhiguang and Wu, Yaoxin and Zhang, Jie},
  title     = {Learning to Handle Complex Constraints for Vehicle Routing Problems},
  booktitle = {Advances in Neural Information Processing Systems},
  year      = {2024},
  editor    = {A. Globerson and L. Mackey and D. Belgrave and A. Fan and U. Paquet and J. Tomczak and C. Zhang},
  volume    = {37},
  pages     = {93479--93509},
  publisher = {Curran Associates, Inc.},
  doi       = {10.52202/079017-2964},
}

@TechReport{Helsgaun2017,
  author      = {Keld Helsgaun},
  title       = {An Extension of the Lin-Kernighan-Helsgaun {TSP} Solver for Constrained Traveling Salesman and Vehicle Routing Problems},
  institution = {Roskilde University},
  year        = {2017},
}

@Article{DashGuenluekLodiTramontani2012,
  author   = {Dash, Sanjeeb and G\"{u}nl\"{u}k, Oktay and Lodi, Andrea and Tramontani, Andrea},
  title    = {A Time Bucket Formulation for the Traveling Salesman Problem with Time Windows},
  journal  = {INFORMS J. Comput.},
  year     = {2012},
  volume   = {24},
  number   = {1},
  pages    = {132-147},
  doi      = {10.1287/ijoc.1100.0432},
  fjournal = {INFORMS Journal on Computing},
}

@Article{ZhengHeZhouJinLi2023,
  author   = {Jiongzhi Zheng and Kun He and Jianrong Zhou and Yan Jin and Chu-Min Li},
  title    = {Reinforced {Lin–Kernighan–Helsgaun} algorithms for the traveling salesman problems},
  journal  = {Knowl. Based Syst.},
  year     = {2023},
  volume   = {260},
  pages    = {110144},
  doi      = {10.1016/j.knosys.2022.110144},
  fjournal = {Knowledge-Based Systems},
}

@Article{CappartMoisanRousseauPremont-SchwarzCire2021,
  author   = {Cappart, Quentin and Moisan, Thierry and Rousseau, Louis-Martin and Prémont-Schwarz, Isabeau and Cire, Andre A.},
  title    = {Combining Reinforcement Learning and Constraint Programming for Combinatorial Optimization},
  journal  = {Proc. AAAI Conf. Artif. Intell.},
  year     = {2021},
  volume   = {35},
  number   = {5},
  pages    = {3677--3687},
  doi      = {10.1609/aaai.v35i5.16484},
  fjournal = {Proceedings of the AAAI Conference on Artificial Intelligence},
}

@Misc{Lopez-IbanezBlum2023,
  author       = {López-Ibáñez, Manuel and Blum, Christian},
  title        = {Benchmark Instances for the Travelling Salesman Problem with Time Windows ({TSPTW})},
  howpublished = {\url{https://lopez-ibanez.eu/tsptw-instances}},
  year         = {2023},
  note         = {(Accessed: 30 November 2025)},
}

\end{document}